\documentclass{aastex62}

\def\ff{f_\mathrm{f}}
\usepackage{ulem}

\shorttitle{Accretion Properties of PDS~70b}
\shortauthors{Hashimoto et al.}

\begin{document}

\title{Accretion Properties of PDS~70b with MUSE\footnote{
    Based on data collected at the European Organisation for Astronomical Research in the Southern Hemisphere under ESO programme 60.A-9100(K). 
    }}

\correspondingauthor{Jun Hashimoto}
\email{jun.hashimto@nao.ac.jp}

\author[0000-0002-3053-3575]{Jun Hashimoto}
\affil{Astrobiology Center, National Institutes of Natural Sciences, 2-21-1 Osawa, Mitaka, Tokyo 181-8588, Japan}
\affil{Subaru Telescope, National Astronomical Observatory of Japan, Mitaka, Tokyo 181-8588, Japan}
\affil{Department of Astronomy, School of Science, Graduate University for Advanced Studies (SOKENDAI), Mitaka, Tokyo 181-8588, Japan}
\author[0000-0003-0568-9225]{Yuhiko Aoyama}
\affil{Department of Earth and Planetary Science, Graduate School of Science, The University of Tokyo, 7-3-1 Hongo, Bunkyo-ku, Tokyo 113-0033, Japan}
\affiliation{Institute for Advanced Study, Tsinghua University, Beijing 100084, People's Republic of China}
\affiliation{Department of Astronomy, Tsinghua University, Beijing 100084, People's Republic of China}
\author[0000-0003-0114-0542]{Mihoko Konishi}
\affil{Faculty of Science and Technology, Oita University, 700 Dannoharu, Oita 870-1192, Japan}
\author[0000-0002-6879-3030]{Taichi Uyama}
\affil{Infrared Processing and Analysis Center, California Institute of Technology, Pasadena, CA 91125, USA}
\affil{NASA Exoplanet Science Institute}
\affil{Subaru Telescope, National Astronomical Observatory of Japan, Mitaka, Tokyo 181-8588, Japan}
\author[0000-0003-3882-3945]{Shinsuke Takasao}
\affil{Division of Science, National Astronomical Observatory of Japan, Mitaka, Tokyo 181-8588, Japan}
\author[0000-0002-5658-5971]{Masahiro Ikoma}
\affil{Department of Earth and Planetary Science, Graduate School of Science, The University of Tokyo, 7-3-1 Hongo, Bunkyo-ku, Tokyo 113-0033, Japan}
\affil{Research Center for the Early Universe (RESCEU), Graduate School of Science, The University of Tokyo, 7-3-1 Hongo, Bunkyo-ku, Tokyo 113-0033, Japan}
\author[0000-0002-5964-1975]{Takayuki Tanigawa}
\affil{National Institute of Technology, Ichinoseki College, Takanashi, Hagisho, Ichinoseki-shi 021-8511, Japan}

\begin{abstract}
We report a new evaluation of the accretion properties of PDS~70b obtained with VLT/MUSE. The main difference from previous studies in \citet{haff19a} and \citet{AI2019} is in the mass accretion rate. Simultaneous multiple line observations, such as H$\alpha$ and H$\beta$, can better constrain the physical properties of an accreting planet. While we clearly detected H$\alpha$ emissions from PDS~70b, no H$\beta$ emissions were detected. We estimate the line flux of H$\beta$ with a 3-$\sigma$ upper limit to be 2.3~$\times$~10$^{-16}$~erg~s$^{-1}$~cm$^{-2}$. The flux ratio $F_{\rm H\beta}$/$F_{\rm H\alpha}$ for PDS~70b is $<$~0.28. Numerical investigations by \citet{aoya18} suggest that $F_{\rm H\beta}$/$F_{\rm H\alpha}$ should be close to unity if the extinction is negligible. We attribute the reduction of the flux ratio to the extinction, and estimate the extinction of H$\alpha$ ($A_{\rm H\alpha}$) for PDS~70b to be $>$~2.0~mag using the interstellar extinction value. 
By combining with the H$\alpha$ linewidth and the dereddening line luminosity of H$\alpha$, 
we derive the PDS~70b mass accretion rate to be $\gtrsim$~5~$\times$~10$^{-7}$~$M_{\rm Jup}$~yr$^{-1}$. 
The PDS~70b mass accretion rate is an order of magnitude larger than that of PDS~70. We found that the filling factor $f_{\rm f}$ (the fractional area of the planetary surface emitting H$\alpha$) is $\gtrsim$0.01, which is similar to the typical stellar value. The small value of $f_{\rm f}$ indicates that the H$\alpha$ emitting areas are localized at the surface of PDS~70b. 
\end{abstract}


\section{Introduction} \label{sec:intro}

Gas giant planets growing in a protoplanetary disk gain their mass via mass accretion from the parent disk until their host star loses its gas disk \citep[e.g.,][]{haya85}. A part of the gas flow from the outer disk accretes onto gas giant planets while the rest of the flow passes over the planets toward the central star. The relationship between the host star and planets in the mass accretion rate depends on the planetary mass and the disk's properties \citep[e.g.,][]{lubow99,tani2016}. Numerical simulations of disk--planet interactions show that the mass accretion rate of a 1-$M_{\rm Jup}$ planet can reach up to $\sim$90~\% of the rate of mass accretion from the outer disk, i.e., the planetary mass accretion rate is an order of magnitude larger than the stellar rate \citep{lubow06}. As predicted by planet population synthesis models \citep[e.g.,][]{ida2004}, when the gas accretion onto planets is sufficiently large, gas giant planets can form more than sub-Jovian planets. On the other hand, microlensing observations show that the number of sub-Jovian planets is dominant in the planetary mass function \citep{suzu2018}. The mass accretion onto a planet determines the final mass of the planet \citep[e.g.,][]{Tanigawa+2007} and thus investigating the planetary mass accretion process informs our understanding of planet formation. 

Direct imaging allows the study of gas accretion in young planets via H$_{\rm I}$ recombination lines such as H$\alpha$ at 656.28~nm \citep[e.g.,][]{zhou14,sall2015,sant18,wagn18a,haff19a}. Although the gas temperature in the surface layers of protoplanetary disks is expected to be high enough ($T \sim$~10$^{4}$~K) to emit H$_{\rm I}$ recombination lines \citep[e.g.,][]{kamp2004}, their emissions may be too weak to be detected due to the low density of the disk surface. Therefore, the observed H$\alpha$ emissions from point-like sources reported in the literature have been attributed to the abundant cold gas around planets in the disk midplane being heated by gas accretion onto planets. In case of T~Tauri stars, two important heating mechanisms in the standard accretion processes have been investigated to understand stellar accretion processes \citep{Bertout+1988,Koenigl1991,poph1993,calv98}. The first is viscous heating in the boundary layer where the faster rotating Keplerian disk connects with the slowly rotating star. The other is accretion shock heating by magnetospheric accretion, in which the gas flow from the midplane of a circumstellar disk shocks the central star by a strong stellar magnetic field. Theoretical studies on planetary accretion processes have also investigated accretion shock heating at the surface of planets \citep[e.g.,][]{zhu2015,tani12,aoya18}.

Recently, \citet{AI2019} demonstrated that the H$\alpha$ spectral line-width and luminosity can be used to constrain the planetary mass and the rate of mass accretion onto protoplanets, by applying radiation-hydrodynamic models of the shock-heated accretion flow \citep{aoya18}. A good example of robustly detected accreting planets at tens of au from a central star is the PDS~70 planetary system. This system includes a weakly accreting young star (spectral type of K7, mass of 0.8~$M_{\odot}$, accretion rate of 6~$\times$~10$^{-11}$~$M_{\odot}$~ yr$^{-1}$, age of 5~Myr, distance of 113~pc; \citealp{peca16,kepp19a,haff19a,mull18a,gaia18}) associated with two planetary-mass companions \citep{haff19a,isell19a,chris19a,chris19b,kepp18a,mull18a,wagn18a}. \citet{AI2019} estimated that PDS~70b's mass and mass accretion rate are $\sim$12~$M_{\rm Jup}$ and $\sim$4~$\times$~10$^{-8}$~$M_{\rm Jup}$~yr$^{-1}$, respectively, while the values for PDS~70c are $\sim$10~$M_{\rm Jup}$ and $\sim$1~$\times$~10$^{-8}$~$M_{\rm Jup}$~yr$^{-1}$. To convert the H$\alpha$ luminosity into the mass accretion rate, we need to constrain the fractional area of the planetary surface emitting H$\alpha$, which is often termed the filling factor $f_{\rm f}$, and the degree of extinction $A_{\rm H\alpha}$ of H$\alpha$. However, these values have been poorly measured in observations.

In this paper, we revisit the accretion properties of PDS~70b by re-analyzing archive data used in \citet{haff19a}. We aim to estimate the physical quantities related to planetary mass accretion with the help of the theoretical study by \citet{AI2019}. We will demonstrate that by combining a 3-$\sigma$ upper limit in the flux ratio of $F_{\rm H\beta}$/$F_{\rm H\alpha}$, the H$\alpha$ line luminosity, and the H$\alpha$ linewidth, we can constrain the filling factor $f_{\rm f}$, the $\rm H\alpha$ extinction $A_{\rm H\alpha}$, and the mass accretion rate onto the planet.

\section{Archival Data and Stellar Spectral Subtraction}

\subsection{Archival Data} \label{sec:data}
Observations of PDS~70 were made with the optical integral field spectrograph MUSE \citep[Multi Unit Spectroscopic Explorer; ][]{bacon10} at the VLT (Very Large Telescope) on June 20th 2018 (UT) under the clear sky conditions with optical seeing of 0.$''$70--0.$''$81, as part of the commissioning runs of the Narrow-Field Mode (NFM). The adaptive optics (AO) system for MUSE consists of the four laser guide star (LGS) facility and the deformable secondary mirror (DSM). The LGS wavefront sensors measure the turbulance in atmosphere, and the DSM corrects a wave-front to provide near diffraction limited images. The data used in this work (a field of view: 7.$''$42~$\times$~7.$''$43, a spatial sampling: 25~$\times$~25~mas$^{2}$, a spectral range: 4650 to 9300~\AA, a resolving power: 2484~$\pm$~5 at 6500.0~\AA) are publicly available on the ESO archive\footnote{\url{http://archive.eso.org/}}. Six individual raw frames with an exposure time of 300 seconds were obtained and calibrated with the MUSE pipeline v2.6 (\citealp{weil2012}; see \citealp{haff19a} for more details about data). The absolute flux was also calibrated in the pipeline where the flux calibration curve is derived from both an atmospheric extinction curve at Cerro Paranal and a spectro-photometric standard star stored as MUSE master calibrations. The apparent magnitude of PDS~70 in calibrated MUSE data at 6905 to 8440~\AA~correspoding to the Sloan $i'$ band filter is measured to 12.0~$\pm$~0.1~mag, which is fainter than the literature value of $i' =$11.1~$\pm$~0.1~mag \citep{hend15}. In the following post-processing, we used an image size of 40~$\times$~40 spatial pixels (corresponding to 1$''$~$\times$~1$''$) around the central star with a high signal-to-noise (SN) ratio. In changing the size, we found that this size maximized the SN of PDS~70b. Furthermore, since there are no signals in $\sim$5780 to $\sim$6050~\AA~to avoid contamination by sodium light from LGS, we removed them from the data.

After the basic calibration with the MUSE pipeline via a scientific workflow of EsoReflex \citep{freu2013}, we extracted six data cubes and inspected them carefully. We found that five of the data cubes have a relatively strong stripe pattern in the vicinity of the central star (Figure~\ref{figA1} in Appendix), induced by a wavelength shift due to possible calibration errors. Since we see no wavelength shift in one data cube, we conducted cross-correlation analyses to correct the wavelength shift. The FWHM of PSF in the final image was 66~mas at 6562.8~\AA.

\subsection{Stellar Spectral Subtraction} \label{sec:pca}
To subtract the stellar spectra of PDS~70 at each spatial pixel, we followed the methods used in \citet{haff19a}. An essential point is to construct reference spectra from the most correlated components obtained from principal component analysis (PCA; see the basis of PCA given by \citealp{jee2007}). 
Before applying PCA, we normalized each spectrum by the median values of each spectrum (a purple line in Figure~\ref{figA3} in Appendix). We then calibrated a fake continuum pattern, as shown in Figure~\ref{figA2} in the Appendix, where we show that the stellar spectra are different from the halo spectra at 300~mas from the stellar position.
This is because the degree of central concentration of the stellar flux is higher at longer wavelengths due to the better AO performance at longer wavelengths.
To calibrate this fake continuum, we generated median spectra for 40~$\times$~40 spatial pixels in six data cubes (a blue line in Figure~\ref{figA3}). We chose the median to avoid residual bad/hot pixels. After dividing each spectrum by the median spectrum (a red line in Figure~\ref{figA3}), we smoothed each divided spectrum by a Gaussian function with a kernel of 230 spectral channels (a black line in Figure~\ref{figA3}). Finally, we divided each normalized spectrum (the purple line in Figure~\ref{figA3}) by the smoothed spectrum (the black line in Figure~\ref{figA3}) to correct the fake continuum (a yellow line in Figure~\ref{figA3}).

We applied PCA to 9,600 spectra (six frames with 40~$\times$~40 spatial pixels) and found the optimum number of components to subtract to be 1, by maximizing the SN of PDS~70b. After subtracting reference spectra for each spatial pixel, we recovered the unit (10$^{-20}$ erg s$^{-1}$ cm$^{-2}$ \AA$^{-1}$) in the image by performing inverse processes of above normalization. Two data cubes out of six were removed from the final image since they degraded the data quality, i.e., gave a worse SN of PDS~70b. The total exposure time after removing the two cubes was 1,200~s. The final images are shown in Figure~\ref{fig1}, in which the H$\alpha$ and H$\beta$ images are constructed with three (656.096, 656.221, and 656.346~nm) and one (486.096~nm) wavelength channels, respectively. Astrometry of the two planets is summarized in Table~\ref{tbl_property}. 

\subsection{Flux Measurement} \label{sec:flux}
\begin{figure*}
    \begin{centering}
    \includegraphics[clip,width=\linewidth]{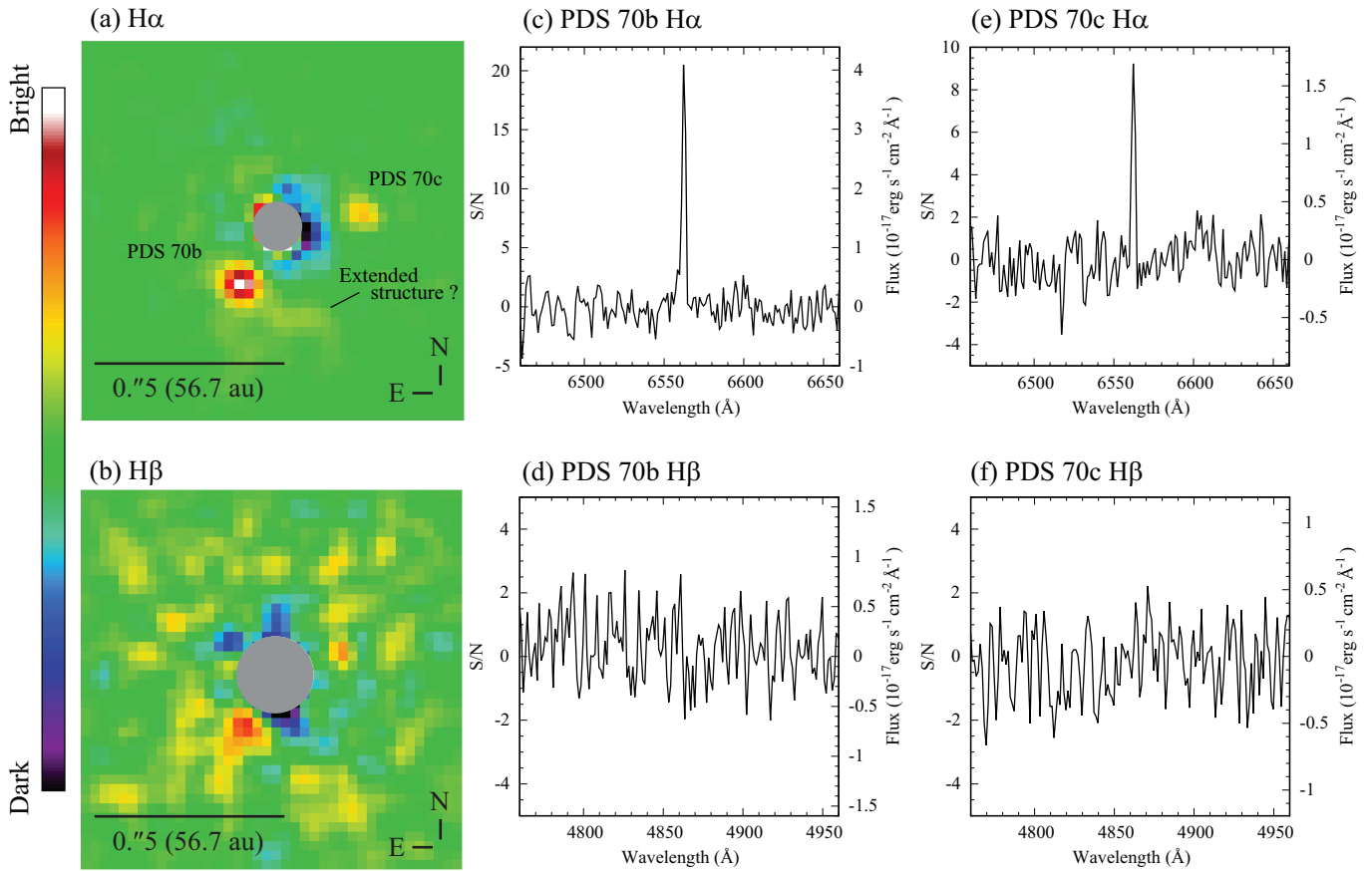}
    \end{centering}
    \caption{H$\alpha$ and H$\beta$ images of PDS~70b and c and their spectra. (a and b) H$\alpha$ and H$\beta$ images generated from three (3.75-\AA\ width) and one (1.25~\AA\ width) spectral channels centered at 6562.21 and 4860.96~\AA, respectively, and are convolved with a Gaussian kernel of 2 pixels in radius. The central regions in panels~(a) and (b) are masked with gray circles of 125 and 200~mas diameter, respectively. (c to f) Spectra of PDS~70b and c at H$\alpha$ and H$\beta$ in a box aperture of 3~$\times$~3 pixels. Neither flux calibration (i.e., aperture and flux correction described in \S~\ref{sec:flux}) nor dereddening correction derived in \S~\ref{sec:extinction} is applied to the spectra.}
    \label{fig1}
\end{figure*}

The H$\alpha$ fluxes were measured with a box aperture of 3~$\times$~3 pixels centered on PDS~70b and c in the combined image of three channels. The errors of H$\alpha$ and H$\beta$ per spectral channel were estimated from the standard deviation of the flux within the box aperture (3~$\times$~3 pixels) at the planetary positions in a continuum spectra, i.e., $\sim$6570 to $\sim$6700~\AA~and $\sim$4800 to $\sim$4930~\AA~for H$\alpha$ and H$\beta$, respectively. The SN ratios of H$\alpha$ were 27 and 10 for PDS~70b and c, respectively (Table~\ref{tbl_property}), while the peak SN ratios along the spectral direction were 20 and 9 for PDS~70b and c, respectively (Figure~\ref{fig1}).

To recover the missing flux, we applied two flux calibrations. The first was an aperture correction \citep{howe90} in which we recovered the missing flux due to the small aperture. The other was a flux correction in which we estimated the lost flux during the PSF subtraction. In the aperture correction, the correction factors (between the box aperture of 3~$\times$~3 pixels and a circular aperture of 70~pixels in radius) derived by photometry of the primary star were 6.65 and 16.9 for H$\alpha$ and H$\beta$, respectively. Since the AO performance is better at longer wavelengths, the correction factor of H$\alpha$ is smaller than that of H$\beta$. We then inspected the extent to which the point source flux decreased during the PSF subtraction. Positive fake sources were injected in the H$\alpha$ and H$\beta$ channels, and the PSF subtraction process was re-run. The fake sources were injected at the separations of PDS~70b and c, and at position angles of 0$\degr$ to 345$\degr$ with 15$\degr$ steps, except the planetary positions, and with fluxes corresponding to those of PDS~70b and c without flux calibrations. Regarding the H$\beta$ flux, we used a 3-$\sigma$ value as the input flux. The results show that the flux decreased to 90\%, 98\%, 90\%, and 95\% for PDS~70b at H$\alpha$, PDS~70c at H$\alpha$, PDS~70b at H$\beta$, and PDS~70c at H$\beta$, respectively. The corrected line fluxes and their errors are shown in Table~\ref{tbl_property}. The linewidths were measured by fitting a Gaussian profile to the H$\alpha$ spectra with the \verb|fit| function in \verb|GNUPLOT|.

\begin{deluxetable}{rcccccc}
    \tablecolumns{1}
    \tablewidth{0pt}
    \tablecaption{Line flux with flux calibration, line profile, and astrometry of PDS 70b and c}\label{tbl_property}
    \tablehead{
        \colhead{Name} & \colhead{Line flux\tablenotemark{a}} &\colhead{Line center\tablenotemark{b}} & \colhead{10\% linewidth\tablenotemark{b}} &  
        \colhead{50\% linewidth} & \colhead{P.A.}& \colhead{Separation} \\
        \colhead{}     & \colhead{(10$^{-16}$~erg~s$^{-1}$~cm$^{-2}$)} & \colhead{(km~s$^{-1}$)} & \colhead{(km~s$^{-1}$)} & 
        \colhead{(km~s$^{-1}$)} & \colhead{($\degr$)}& \colhead{(mas; au)}    
    } 
    \startdata
        PDS~70b (H$\alpha$) & 8.1~$\pm$~0.3 & $-$18 $\pm$ 4 & 213~$\pm$~15 & 114~$\pm$~8 & 147~$\pm$~8 & 178~$\pm$~25; 20.2~$\pm$~2.8 \\
                 (H$\beta$)  & $<$2.3\tablenotemark{c}& ---& --- & --- & --- \\
        PDS~70c (H$\alpha$) & 3.1~$\pm$~0.3 & $-$26 $\pm$ 6 & 200~$\pm$~24 & 107~$\pm$~13 & 278~$\pm$~6 & 225~$\pm$~25; 25.5~$\pm$~2.8 \\
                 (H$\beta$)  & $<$1.6\tablenotemark{c} & --- & --- & --- & --- \\
    \enddata

    \tablenotetext{a}{Flux calibration (i.e., aperture and a flux correction described in \S~\ref{sec:flux}) is applied to values, while values are not dereddened with $A_{H\alpha}$ and $A_{H\beta}$ derived in \S~\ref{sec:extinction}. Dereddened H$\alpha$ fluxes are $>$50.6 and $>$8.5~$\times$~10$^{-16}$~erg~s$^{-1}$~cm$^{-2}$ for PDS~70b and c, respectively (see details in \S~\ref{sec:extinction}).}
    \tablenotetext{b}{Line center relative to the rest wavelength of H$\alpha$ at 6562.8~\AA~is measured by Gaussian fitting. For PDS~70, the line center and the 10\% \& 50\% linewidths are $-$28~$\pm$~7, 283~$\pm$~28~km~s$^{-1}$, 151~$\pm$~15~km~s$^{-1}$, respectively.}
    \tablenotetext{c}{3~$\sigma$ upper limit.}
\end{deluxetable}

\section{Results}\label{sec:result}

\begin{figure*}
    \begin{centering}
    \includegraphics[clip,width=\linewidth]{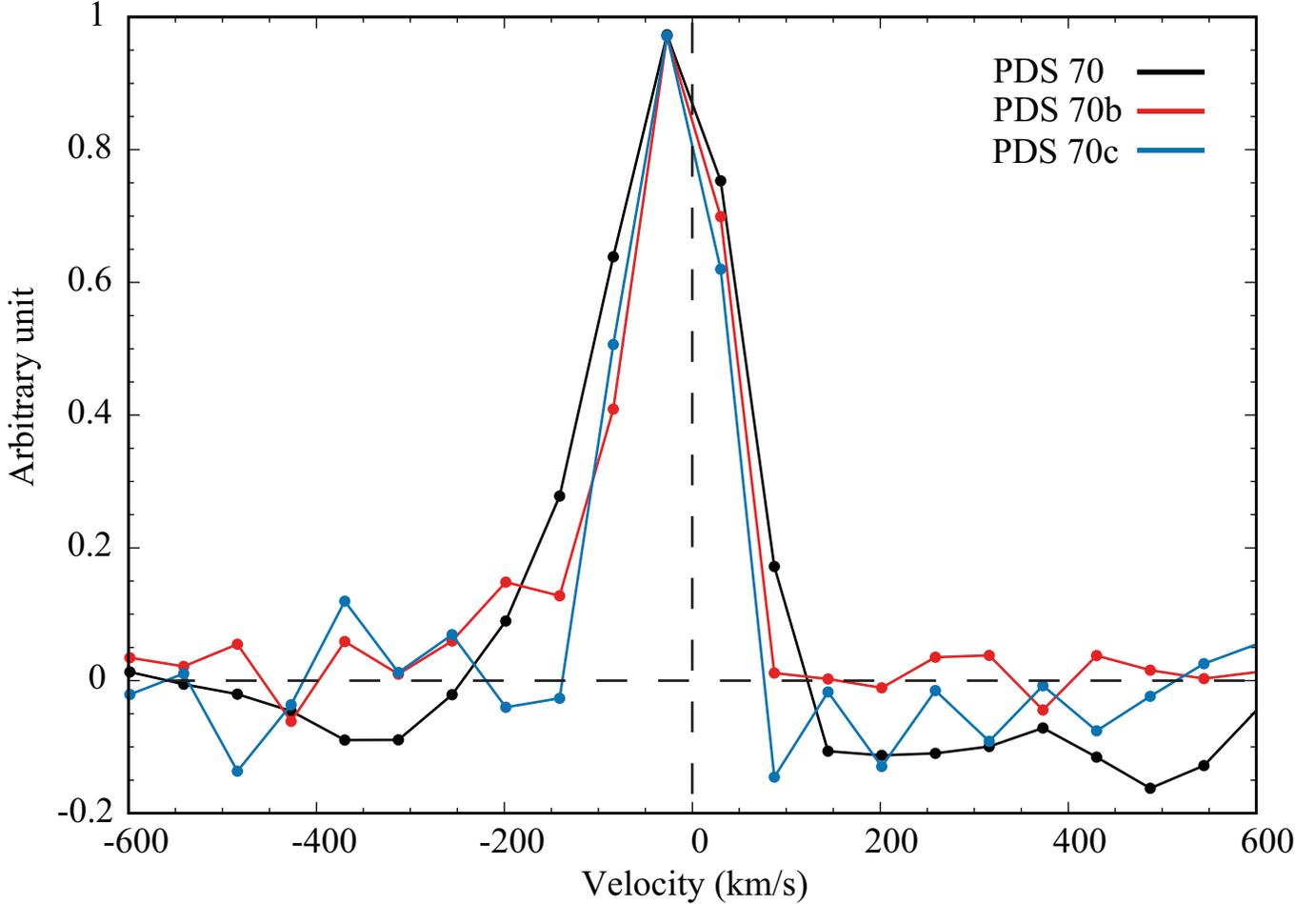}
    \end{centering}
    \caption{Normalized H$\alpha$ line profiles of PDS~70 and its two planets. The planetary H$\alpha$ data are the same as those in Figure~\ref{fig1}, but are normalized. The vertical dotted line represents the rest wavelength of H$\alpha$ at 6562.8~\AA.}
    \label{fig2}
\end{figure*}

Figure~\ref{fig1} and Table~\ref{tbl_property} show our re-analysis results. They are largely similar with those in \citet{haff19a}. As already reported in \citet{haff19a}, we also clearly detected H$\alpha$ emissions from PDS~70b and c, while H$\beta$ emissions were not detected. The peak SN ratio of $\sim$20 in PDS~70b (Figure~\ref{fig1}c) is roughly 2$\times$ better than that in \citet{haff19a}. The H$\alpha$ line fluxes with flux calibration described in \S~\ref{sec:flux} are 8.1~$\pm$~0.3~$\times$10$^{-16}$ and 3.1~$\pm$~0.3~$\times$10$^{-16}$~erg~s$^{-1}$~cm$^{-2}$ for PDS~70b and c, respectively. The 50~\% linewidths (FWHM) of $\sim$110~km~s$^{-1}$ in the two planets are comparable with a spectral resolution of $\sim$120~km~s$^{-1}$ in MUSE, and thus, these values could be upper limits. The main differences in our results from those in \citet{haff19a} are the H$\alpha$ line profiles.
Figure~\ref{fig2} shows the normalized H$\alpha$ line profiles of the primary star, PDS~70b, and PDS~70c. All three spectra exhibit a blue-shifted single Gaussian profile at the line center of $-$20 to $-$30~km~s$^{-1}$ ($-$0.4 to $-$0.7~\AA). 
According to the MUSE manual, this blue shift could be due to wavelength calibration errors, as the wavelength calibration errors are more than 0.4~\AA~based on only the arc frames taken in the morning calibrations. Note that the radial velocity of PDS~70 ($\sim$3~km~s$^{-1}$: 0.07~\AA; \citealp{gaia18}) and the Keplerian velocities of PDS~70b and c ($\sim$3--4~km~s$^{-1}$: 0.07--0.09~\AA) are negligible.

We found a possible extended structure in the south west of PDS~70b in Figure~\ref{fig1}(a). This structure was also reported in \citet{haff19a}. Although the SN ratio for this structure is $\sim$3--4, the structure is exactly located at the image slicing and field splitting axis in one data cube (see Figure~\ref{figA1} in Appendix). Therefore, \citet{haff19a} regarded this structure as an artifact. However, as described in \S~\ref{sec:pca}, since we removed this data cube from the final image to improve the SN of PDS~70b, this structure might be marginally real. Since the study of the nature of this structure is beyond the scope of this paper, the investigations are left for future work. 
Furthermore, \citet{mesa2019} recently reported the detection of a point-like feature (PLF) at $r \sim$~0.$''$12, while no significant H$\alpha$ emission can be seen in our results.

\section{Physical Properties of Accreting Planets}\label{sec:model}
\subsection{Free-fall Velocity of Gas and Pre-shock Number Density of Hydrogen Nuclei}
\begin{figure}
    \centering
    \includegraphics[clip,width=\linewidth]{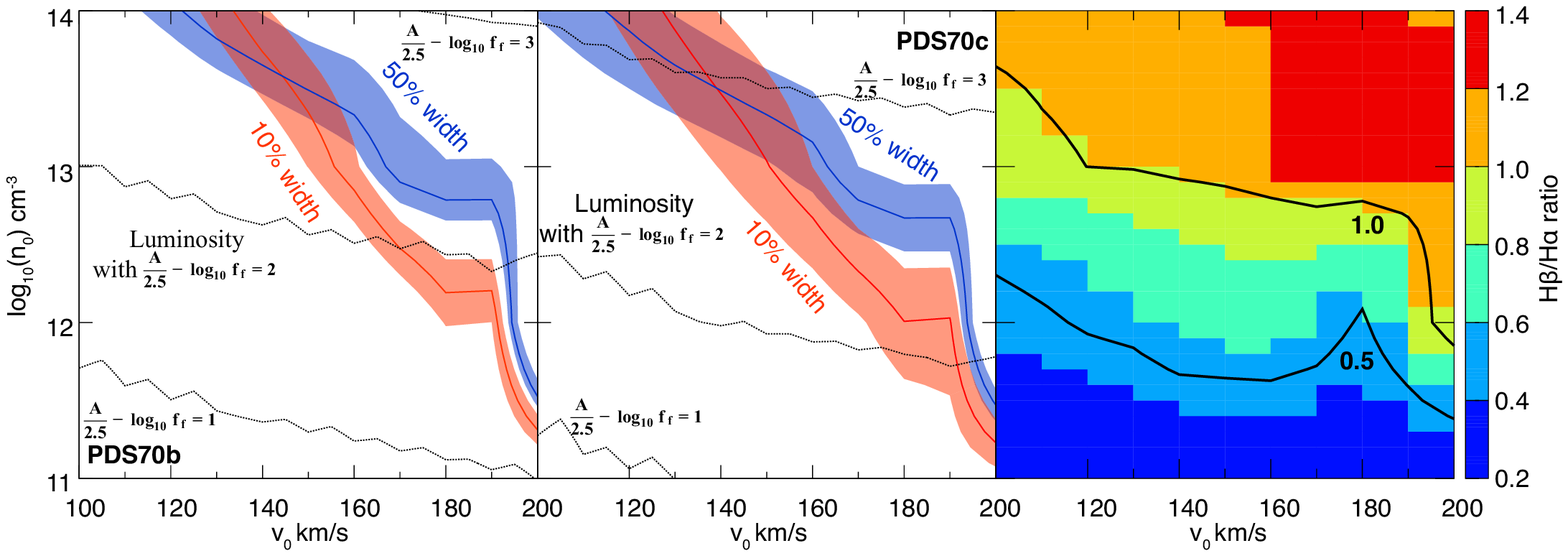}
    \caption{Model input parameters $v_0\,\mathrm{km/s}$ and $n_0\,\mathrm{cm^{-3}}$ reproducing the H$\alpha$ line profile observed in PDS~70b (left) and c (center). The red and blue lines correspond to the H$\alpha$ linewidth at 10~\% and 50~\% of maximum flux, respectively. The colored shades show their 1~$\sigma$ errors taken from fitting errors in Table~\ref{tbl_property}.
    The black dotted lines show the dereddened line luminosity for the planetary radius of $2R_\mathrm{J}$ and different filling factors $f_\mathrm{f}$. The cross points of the red and blue lines roughly indicate $(v_0,~n_0)=(144~\mathrm{km/s},~3.8\times10^{13}~\mathrm{cm^{-3}})$ and $(139~\mathrm{km/s},~3.2\times10^{13}~\mathrm{cm^{-6}})$ for PDS70b and c, respectively. To reproduce the dereddened luminosity at these cross points, the values of $10^{A/2.5} / f_\mathrm{f}$ are estimated to be $5\times 10^{2}$ and $9\times10^{2}$, respectively. Note that the values of $A_{H\alpha}$ used here are 2.0 and 1.1~mag for PDS~70b and c, respectively (\S~\ref{sec:extinction}). 
    The right panel shows the ratio of H$\beta$ to H$\alpha$. The cross points in the left and middle panels indicate that the flux ratio is close to unity for PDS~70b and c.} 
    \label{fig3}
\end{figure}

The simultaneous observations of H$\alpha$ and H$\beta$ with MUSE provide reliable measurements of the line flux ratio without any concerns regarding the temporal variability of each flux. Signals of H$\beta$ emissions are not seen in both PDS~70b and c, and we only obtained the upper limit of its flux. However, the upper limit of H$\beta$ emissions allows us to constrain the physical parameters related to the planetary accretion.  
Following the method in \citet{AI2019}, we first estimated the free-fall velocity $v_0$ of the gas and the pre-shock number density of hydrogen nuclei $n_0$ using the accretion shock model in \citet{aoya18}. Here, we assume that the accretion speed is the same as the free-fall velocity. 
The observational constraints for these quantities are given by the H$\alpha$ linewidths at 10~\% and 50~\% maxima (Table~\ref{tbl_property}). As explained in \citet{AI2019}, the H$\alpha$ linewidth increases with $n_0$ and $v_0$. This is because an increase of $n_0$ enhances absorption of H$\alpha$ emissions near the line center by shock-heated gas, while a high value of $v_0$ means that a flow with higher velocity passes. Thus the H$\alpha$ linewidth is a function of $n_0$ and $v_0$ \citep[see Figure~5 in][]{AI2019}. In Figure~\ref{fig3}, the intersections of the H$\alpha$ 10~\% \& 50~\% linewidths indicate that ($v_0$, $n_0$) is roughly ($144$~km~s$^{-1}$, $3.8\times10^{13}$~cm$^{-3}$) and ($139$~km~s$^{-1}$, $3.2\times10^{13}$~cm$^{-3}$) for PDS~70b and c, respectively. 

\subsection{Line Flux Ratio and Extinction}\label{sec:extinction}
The theoretical model of \citet{AI2019} suggests that the flux ratio of H$\beta$ to H$\alpha$ ($F_{\rm H\beta}/F_{\rm H\alpha}$) increases with $n_0$ when there is no foreground extinction, e.g., interstellar extinction (right panel in Figure~\ref{fig3}). This is because the higher density of $n_{0} \gtrsim$~10$^{13}$~cm$^{-3}$ is large enough to saturate the H$\alpha$ flux due to the effect of absorption in the post-shock region. Meanwhile, since the emission source for H$\beta$ is more excited than that for H$\alpha$, higher values of $v_0$ or $n_0$ are necessary to put H$\beta$ in such a saturated state (see \citealp{aoya18} for more details). Hence the value of $F_{\rm H\beta}$/$F_{\rm H\alpha}$ is close to unity for $n_{0} \gtrsim$~10$^{13}$~cm$^{-3}$. In contrast, a H$\alpha$ absorption is negligible with the lower density of $n_{0} \lesssim$~10$^{12}$~cm$^{-3}$, resulting in a smaller value of $F_{\rm H\beta}/F_{\rm H\alpha} \lesssim$~0.5.

With values of $n_0$ and $v_0$ as derived above for PDS~70b and c, the theoretical value of the flux ratio is close to unity (right panel in Figure~\ref{fig3}). However, our observed values of $F_{\rm H\beta}$/$F_{\rm H\alpha}$ at a 3-$\sigma$ upper limit for PDS~70b and c are $<$~0.28 and $<$~0.52, respectively, which are smaller than the theoretically expected values. These observed values are also similar with accreting T~Tauri stars and brown dwarfs with the values of 0.1--0.5 \citep[see Table~7 in][]{herc08}. This underestimating of the flux ratio could be due to extinction of circumplanetary materials. According to ALMA observations and numerical simulations by \citet{kepp19a}, gas is likely to exist in the dust gap of the circumstellar disk where PDS~70b and c are located. We conjecture that small dust grains (sub-micron size) are well coupled with gas in the gap, contributing to the extinction. Based on this speculation, we use the extinction law $A_{\lambda} \propto \lambda^{-1.75}$ \citep{draine89} to estimate the extinction. Considering the theoretical prediction that the value of $F_{\rm H\beta}$/$F_{\rm H\alpha}$ without extinction is unity, we derived $A_{H\alpha} >$~2.0 and $>$~1.1~mag for PDS~70b and c, respectively. We note that the extinction law is valid in the range $0.7 \lesssim \lambda \lesssim 5 \mu$m, and the slope of extinction will be gentle at H$\alpha$ and H$\beta$ wavelengths (see Figure~1 in \citealp{draine89}). The gentle slope will make the lower limit of $A_{H\alpha}$ larger than the above value.

\subsection{Filling Factor, Planetary Mass, and Mass Accretion Rate}\label{sec:ff}
We estimate the filling factor $\ff$ using the following equation for the H$\alpha$ luminosity \citep{AI2019}: 
\begin{equation}
L_{\rm H_\alpha} = 4\pi R_{\rm p}^2 \ff I_{\rm H_\alpha} 10^{-0.4A_{\rm H_\alpha}},\label{eq:LHa}
\end{equation}
or 
\begin{equation}
0.4A_{\rm H_\alpha} - \log_{10}\ff = -\log_{10}L_{\rm H_\alpha} + \log_{10}(4\pi R_{\rm p}^2 I_{\rm H_\alpha}),\label{eq:LHa2}
\end{equation}
where $R_{\rm p}$ and $I_{\rm H_\alpha}$ are the planetary radius and the H$\alpha$ energy flux per unit area, respectively. The H$\alpha$ luminosities are calculated from the line fluxes listed in Table~\ref{tbl_property}. We assume that the planetary radius $R_{\rm p}$ is $2R_{\rm Jup}$, where $R_{\rm Jup}$ is the Jovian radius \citep{spie12}. $I_{\rm H_\alpha}$ is a function of $n_0$ and $v_0$, and its dependence is investigated in \citet{aoya18}. Using these results, we plot the lines of $0.4A_{\rm H_\alpha} - \log_{10}\ff$ with different values in the $v_0$--$n_0$ plane (left and middle panels in Figure~\ref{fig3}). The lower limits for the extinction $A_{H\alpha}$ constrain the possible range of the filling factor: $A_{H\alpha} >$~2.0 and $>$~1.1~mag give $\ff\gtrsim 0.01$ and $\gtrsim0.003$ for PDS~70b and c, respectively. 

We can also derive a planetary dynamical mass $M_\mathrm{P}$ and an accretion rate $\dot{M}_\mathrm{P}$ \citep{AI2019}, written as
\begin{eqnarray}
    M_\mathrm{P} &=& \frac{R_\mathrm{P}v_0^2}{2G} ~ \mathrm{and}\label{eq:Mp}\\
    \dot{M}_\mathrm{P} &=& \mu n_0 v_0 \ff (4\pi R_\mathrm{P}^2),\label{eq:Mp_dot}
\end{eqnarray}
respectively, where $G$ is the gravitational constant and $\mu$ is the mean mass per hydrogen nucleus. Using $\mu=2.3\times10^{-24}$\,g and $R_{\rm P} =$~2~$R_{\rm Jup}$, we estimate $M_\mathrm{P}$ of 12~$\pm$~3~$M_{\rm Jup}$ and $\dot{M}_\mathrm{P}\gtrsim$~5$\times10^{-7}\,M_\mathrm{Jup}\,\mathrm{yr^{-1}}$ for PDS~70b, and
$M_\mathrm{P}$ of 11~$\pm$~5~$M_{\rm Jup}$ and $\dot{M}_\mathrm{P}\gtrsim$~1$\times10^{-7}\,M_\mathrm{Jup}\,\mathrm{yr^{-1}}$ for PDS~70c.

\section{Discussion}\label{sec:discuss}

\subsection{Planetary Mass}\label{sec:mass}
The planetary mass is commonly estimated by comparing photometric and spectroscopic observations with planetary evolution and atmospheric models \citep[e.g.,][]{mull18a,kepp18a,chris19a,haff19a,mesa2019}. On the other hand, our method based on the accretion shock model can estimate a dynamic planetary mass \citep{AI2019}. In \S~\ref{sec:ff}, we estimated the masses of PDS~70b and c to be $\sim$12 and $\sim$11~$M_{\rm Jup}$, respectively. As shown in eq.~\ref{eq:Mp}, these values depend on the planetary radii, which cannot be constrained only by MUSE observations. We consider the possible range of the planetary radii suggested by the planetary evolution model, which predicts the radii to be $\sim$1--2~$R_{\rm Jup}$ with an age of $\lesssim 10$~Myr \citep{spie12}. Therefore, the planetary mass derived from the radii of $2R_{\rm Jup}$ will give the upper limit. We also note that the observed linewidths will be the upper limits, because the linewidths of the two planets are similar to MUSE's spectral resolution, $\sim$120~km~s$^{-1}$. Our estimated masses are consistent with previous near-infrared photometric and spectroscopic estimations: the mass of PDS~70b is estimated to be 5--9 and 2--17~$M_{\rm Jup}$ from photometric and spectroscopic observations, respectively \citep{kepp18a,mull18a}, while PDS~70c's mass is 4--12~$M_{\rm Jup}$ from photometric observations \citep{haff19a,mesa2019}. Our mass estimation method is independent of the previous methods based on the planetary evolution and atmospheric models only, and therefore will provide a powerful tool to calibrate these models. 

\subsection{Origin of Planetary Extinction}
The extinction of the primary star was found to be negligible \citep{mull18a}. However, our observations suggest that PDS~70b and c have $A_{\rm H \alpha} >$~2.0 and $>$~1.1~mag, respectively. We hypothesize that small 
(sub-micron size) grains coupled with the gas in the dust gap in the circumstellar disk is the cause of the planetary extinction. Such dust, if it exists, will not be clearly visible in existing observations because of their faintness. Here, we estimate the contribution of the unseen dust to the extinction. \citet{kepp19a} conducted hydrodynamic simulations to reproduce the observed $^{12}$CO (2--1) integrated intensity map and found that the azimuthally averaged gas surface density at the location of PDS~70b is $\sim$0.01~g~cm$^{-2}$, which can be translated to a molecular hydrogen surface density of $\sim$3~$\times$~10$^{21}$~cm$^{-2}$. By using the relationship between the interstellar dust extinction and the hydrogen column density ($N_{\rm H}$/$A_{V} = 1.9 \times 10^{21}$~cm$^{-2}$~mag$^{-1}$; \citealp{bohlin78}), we derived $A_{V} \sim$~3.3~mag, which corresponds to $A_{H\alpha}$ $\sim$~2.4~mag with $A_{\lambda} \propto \lambda^{-1.75}$ \citep{draine89}, where we assume that hydrogen atoms are in the form of hydrogen molecules. The derived vertical extinction at the gap region is comparable to the planetary extinction estimated by the accretion shock model, which supports our idea that the origin of extinction is small, unseen dust grains in the gap.

\subsection{Planetary Mass Accretion Rate}\label{sec:mdot}

We now consider the influence of the extinction on the estimation of the mass accretion rate. First, we briefly explain the relationship between the extinction and the mass accretion rate. As shown in eq.~\ref{eq:Mp_dot}, the mass accretion rate is a function of $\ff$, $n_{0}$, and $v_{0}$. The values of $n_{0}$ and $v_{0}$ can be estimated from the H$\alpha$ linewidths (left and middle panels in Figure~\ref{fig3}), and hence we only need to estimate the value of $\ff$. This value can be estimated from eq.~\ref{eq:LHa}, with the free parameter $A_{\rm H \alpha}$. The value of $A_{\rm H \alpha}$ was independently estimated from the line ratio (right panel in Figure~\ref{fig3}). For PDS~70b, we derived an H$\alpha$ extinction of $A_{\rm H \alpha} >$~2.0~mag in the pre-shock region (\S~\ref{sec:extinction}). With eq.~\ref{eq:LHa} and \ref{eq:Mp_dot}, the value of the mass accretion rate derived from the dereddened luminosity is $\dot{M}_\mathrm{P}\gtrsim$~5~$\times$~10$^{-7}$~$M_{\rm Jup}$~yr$^{-1}$ (left panel in Figure~\ref{fig3}). The PDS~70c mass accretion rate was revised to $\gtrsim$~1~$\times$~10$^{-7}$~$M_{\rm Jup}$~yr$^{-1}$ with $A_{\rm H \alpha} >$~1.1~mag (middle panel in Figure~\ref{fig3}). Note that our estimation of the mass accretion rate in this paper is limited by the detection limit of the H$\beta$ flux, and the true mass accretion rate should be higher than the current values. In particular, we speculate that the intrinsic mass accretion rate of PDS~70c is larger than that of PDS~70b because the infrared color of $K1$--$L'$ in PDS~70c with $\sim$2.2~mag is redder than that of PDS~70b with $\sim$1.1~mag \citep{haff19a}, resulting in a much larger value of $A_{\rm H\alpha}$ for PDS~70c. To better constrain the mass accretion rate of PDS~70b and c, deep multiple-line observations with less extinction, such as Pa$\beta$ (1.282~$\mu$m) and Br$\gamma$ (2.166~$\mu$m), are preferable.

Our analysis suggests that the mass accretion rates of PDS~70b and c are $\gtrsim$~8$\times$ and $\gtrsim$~2$\times$ higher than that of the primary star ($\sim$6~$\times$~10$^{-8}$~$M_{\rm Jup}$~yr$^{-1}$; \citealp{haff19a}). Recently, \citet{than2020} applied magnetospheric accretion models to the H$\alpha$ line profile of PDS~70 and derived mass accretion rates onto the star in the range of 0.6--2.2~$\times$~10$^{-7}$~$M_{\rm Jup}$~yr$^{-1}$. Even with these new stellar values, the mass accretion rate of PDS~70b is still higher. Since the number of accreting planets embedded in the protoplanetary disk is currently believed to be three (LkCa~15b, PDS~70b, and PDS~70c; \citealp{kraus2012,sall2015,wagn18a,haff19a}), it is uncertain whether this situation (i.e., a higher planetary accretion rate) is common or rare. Numerical simulations predict that the mass accretion rate of a 1-$M_{\rm Jup}$ planet can reach up to $\sim$90~\% of the gas flow from the outer disk \citep{lubow06}. Furthermore, \citet{tani2016} showed that the mass accretion rate of a planet in the gas gap of the disk can exceed the stellar mass accretion rate in the case of a lower planetary mass and/or a higher gas scale height (see eq.~16 in \citealp{tani2016}). Therefore, a situation similar to the PDS~70 system can occur at a certain evolution phase in other disk systems. 

Additionally we compared our results with the mass accretion rates of other young planetary-mass companions \citep[GSC~06214--0210b, GQ~Lup~b, DH~Tau~b, and SR~12c in ][]{bowl2011,zhou14,sant18}. The mass accretion rate of GQ~Lup~b ($\dot{M}_\mathrm{P}\sim$~5~$\times$~10$^{-7}$~$M_{\rm Jup}$~yr$^{-1}$; \citealp{zhou14}) is comparable with that of PDS~70b, while other three objects have lower values by a few orders of magnitude ($\dot{M}_\mathrm{P}\sim$~10$^{-9}$--10$^{-8}$~$M_{\rm Jup}$~yr$^{-1}$; \citealp{bowl2011,zhou14,sant18}). These results of the higher mass accretion rates of PDS~70b and GQ~Lup~b could be naturally explained by the fact that these two are embedded in the circumstellar disk \citep{kepp19a,macg2017} and can be supplied with a fresh disk material from the parent disk.

\subsection{Accretion Process}\label{sec:accretion}

Our analysis based on the accretion shock model suggests that $\ff\sim10^{-2}$--10$^{-3}$ for PDS~70b and c. If the H$\alpha$ emitting regions are actually accretion shock regions, the accretion flow toward the protoplanets should be significantly converged by some mechanisms. The existence of accretion shocks by converging flows has been suggested for pre-main-sequence stars such as classical T-Tauri stars, where strong stellar magnetic fields guide and collimate the accretion flows toward the magnetic poles \citep{Koenigl1991,hart16}. The very small filling factors for PDS~70b and c may imply that magnetospheric accretion operates even in these protoplanets \citep[for recent theoretical studies, see][]{baty18,than2019}. However, the existence of sufficiently strong planetary magnetic fields remains unclear, and needs to be examined observationally \citep[e.g., radio observations with a low frequency of 10s of MHz are currently being performed:][]{zark19}. If there are no magnetic fields or only a weak field in PDS~70b and c, some hydrodynamic processes should be responsible for collimating the
accretion flows.

When a protoplanet does not develop a magnetosphere due to it having a weak magnetic field, and its circumplanetary disk extends to the planetary surface, the boundary layer between the protoplanet and the disk is heated due to a viscous process and can be a source of H$\alpha$ emissions. Boundary layer accretion has also been considered for pre-main-sequence stars \citep{Bertout+1988}.
Further theoretical and observational studies are required to identify the accretion processes. For example, high spectral resolution observations to search for the inverse P~Cygni profile will help to investigate the possibility of magnetospheric accretion \citep[e.g., AA~Tau; ][]{edwa1994}. Observational investigations of planetary magnetic fields at radio wavelengths will give constraints on the field strength \citep[e.g., SKA: the Square Kilometre Array;][]{zark15}. Multi-epoch observations of accreting signatures at optical/NIR wavelengths (e.g., VLT/MUSE or Keck/OSIRIS) will provide the information on time-variability; magnetospheric accretion in pre-main-sequence stars commonly show time-variability \citep[e.g.,][]{cody2014}.

\section{Summary}\label{sec:summary}
We re-analyzed MUSE archive data obtained with commissioning observations, and estimated the upper limit of H$\beta$ emissions for PDS~70b. Most of the observational results are similar to those of previous studies in \citet{haff19a}. The main difference is the planetary mass accretion rate. We derived the accretion properties of PDS~70b by assuming that the H$\alpha$ emissions originate from gas accretion shock. We showed that the line flux ratio $F_{\rm H\beta}$/$F_{\rm H\alpha}$ is useful to constrain the planetary mass accretion rate by estimating the extinction of $A_{H\alpha}$ because the accretion rate is described as a function of $A_{H\alpha}$. The 3-$\sigma$ upper limit of $F_{\rm H\beta}$/$F_{\rm H\alpha} =$~0.28 can be translated to $A_{H\alpha} >$~2.0~mag. We then obtained a value for dereddened $F_{\rm H\alpha} >$~5~$\times$~10$^{-15}$~erg~s$^{-1}$~cm$^{-2}$. With the H$\alpha$ linewidth and the dereddened H$\alpha$ line luminosity for PDS~70b, we derived a mass accretion rate of $\dot{M} \gtrsim$~5~$\times$~10$^{-7}$~$M_{\rm Jup}$~yr$^{-1}$ for PDS~70b. PDS~70b's mass accretion rate is an order of magnitude larger than that of PDS~70 with $\dot{M} \sim$~6~$\times$~10$^{-8}$~$M_{\rm Jup}$~yr$^{-1}$. We also derived a filling factor $f_{\rm f}$ of $\gtrsim$~0.01 for PDS~70b. This result suggests that the H$\alpha$ emitting areas are localized at the surface of PDS~70b. Multiple line observations, especially emission lines with low extinction, such as Pa$\beta$ (1.282~$\mu$m) and Pa$\delta$ (2.166~$\mu$m), are useful for determining better constraints on the planetary accretion properties of young planets deeply embedded in molecular clouds or circumstellar disks.

\acknowledgments
We thanks an anonymous referee for a helpful review of the manuscript.
We are grateful to K.~Kanagawa and Y.~Hasegawa for their contributions.
This work was supported by JSPS KAKENHI Grant Numbers 19H00703, 19H05089, 19K03932, 18K13579, and 15H02065.

\appendix

\begin{figure*}[tbh!]
    \begin{centering}
    \includegraphics[clip,width=\linewidth]{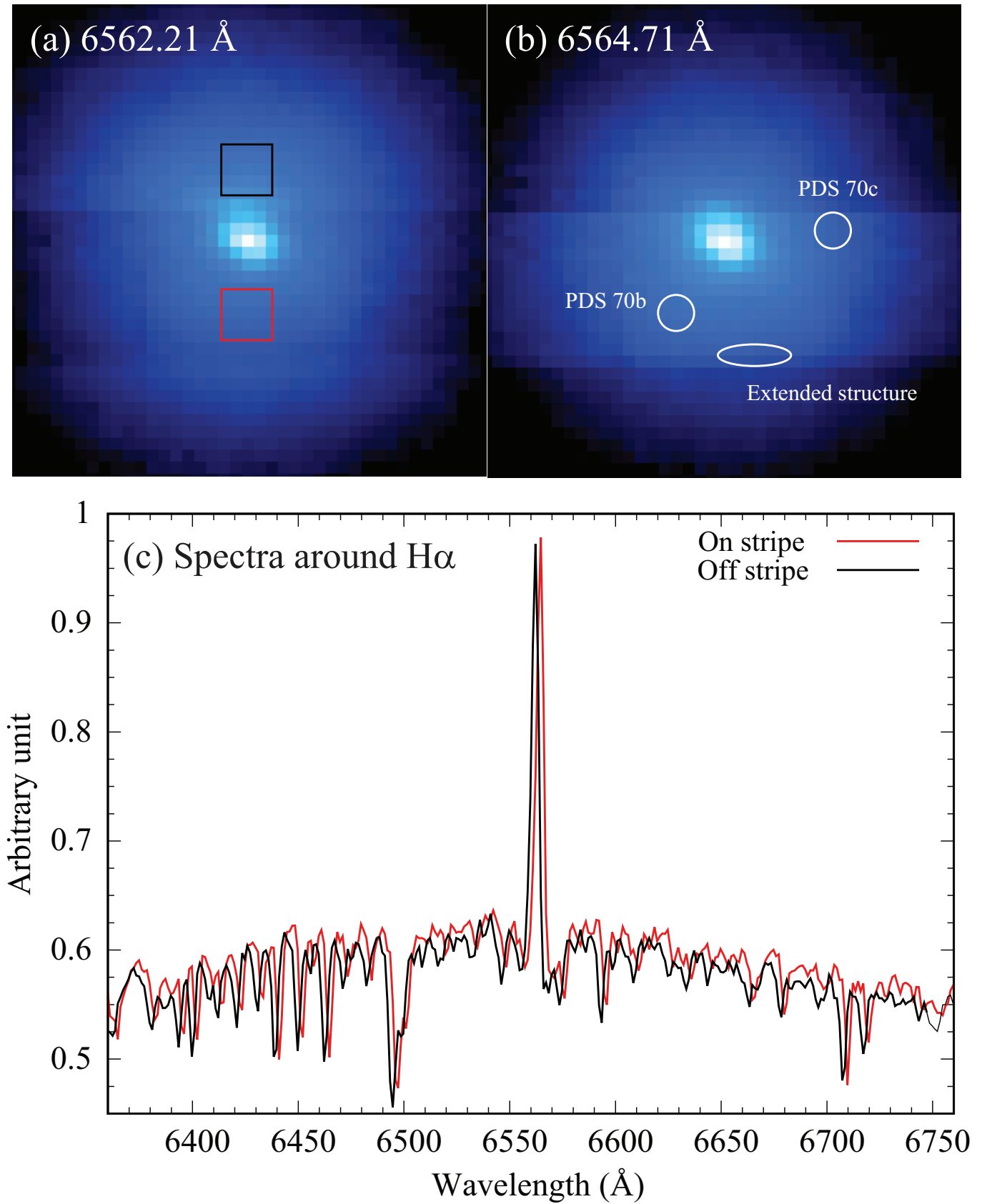}
    \end{centering}
    \caption{Stripe patterns in the vicinity of the central star at 6562.21~\AA~in panel (a) and 6564.71~\AA~in panel (b). The image size is 1$''$~$\times$~1$''$. The white circles and ellipse indicate the positions of PDS~70b, PDS~70c, and a possible extended structure. (c) Spectra in the stripe (red) and in the off-stripe (black) at the red and black square regions in panel (a), respectively. The stripe pattern is induced by a wavelength shift possibly due to calibration errors.}
    \label{figA1}
\end{figure*}

Figure~\ref{figA1} shows wavelength shifts in a stripe pattern, which is corrected in \S~\ref{sec:data}.

\begin{figure*}
    \begin{centering}
    \includegraphics[clip,width=\linewidth]{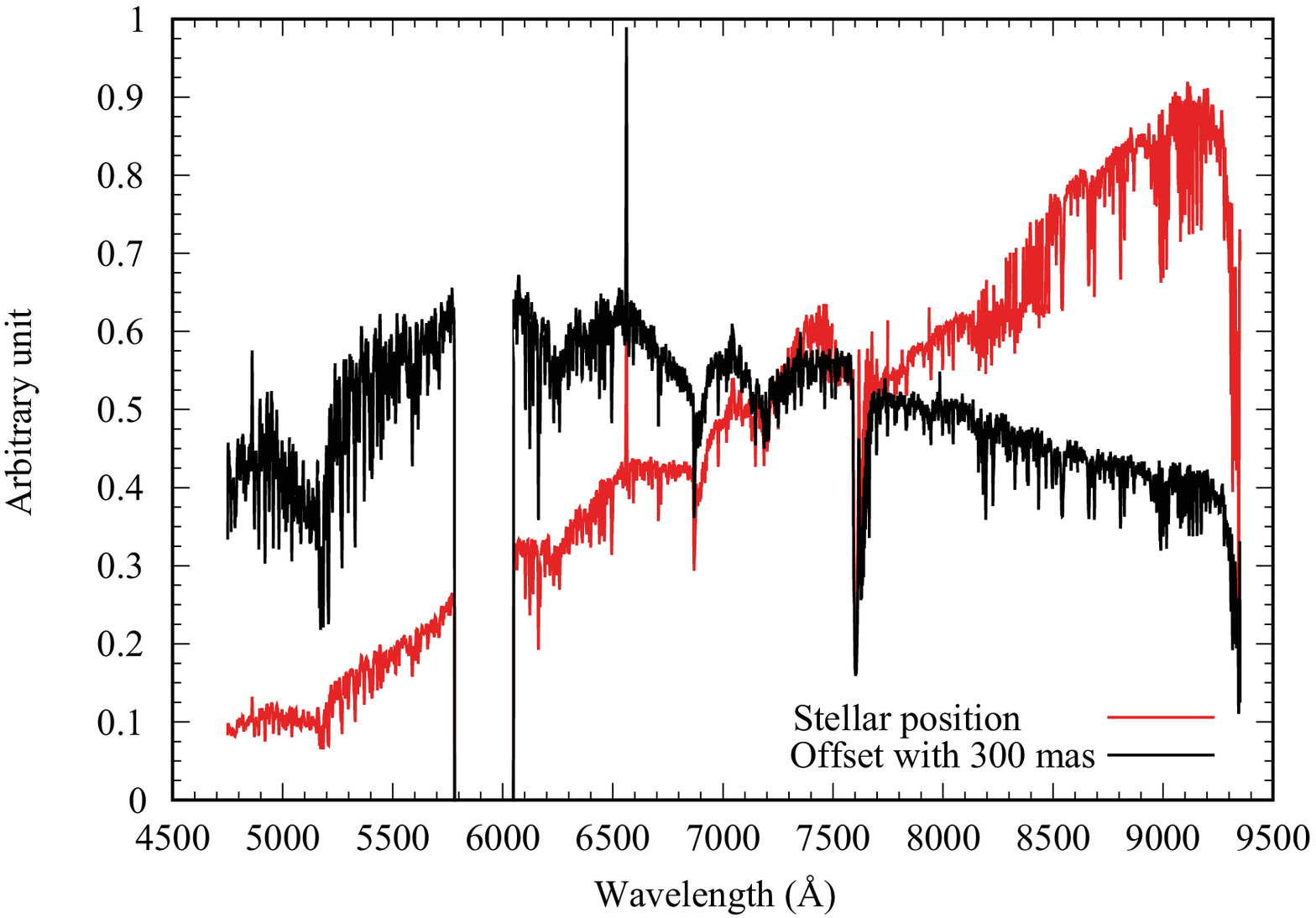}
    \end{centering}
    \caption{Spectra in the 3~$\times$~3~pixels area at the stellar position (red) and at the halo region with a 300-mas offset from the stellar position (black). Since adaptive optics systems work well at longer wavelengths, the flux in the arbitrary box at the stellar position is brighter at longer wavelengths.}
    \label{figA2}
\end{figure*}

Figure~\ref{figA2} shows different spectra in the stellar and halo regions, which is corrected in \S~\ref{sec:pca}.

\begin{figure*}
    \begin{centering}
    \includegraphics[clip,width=\linewidth]{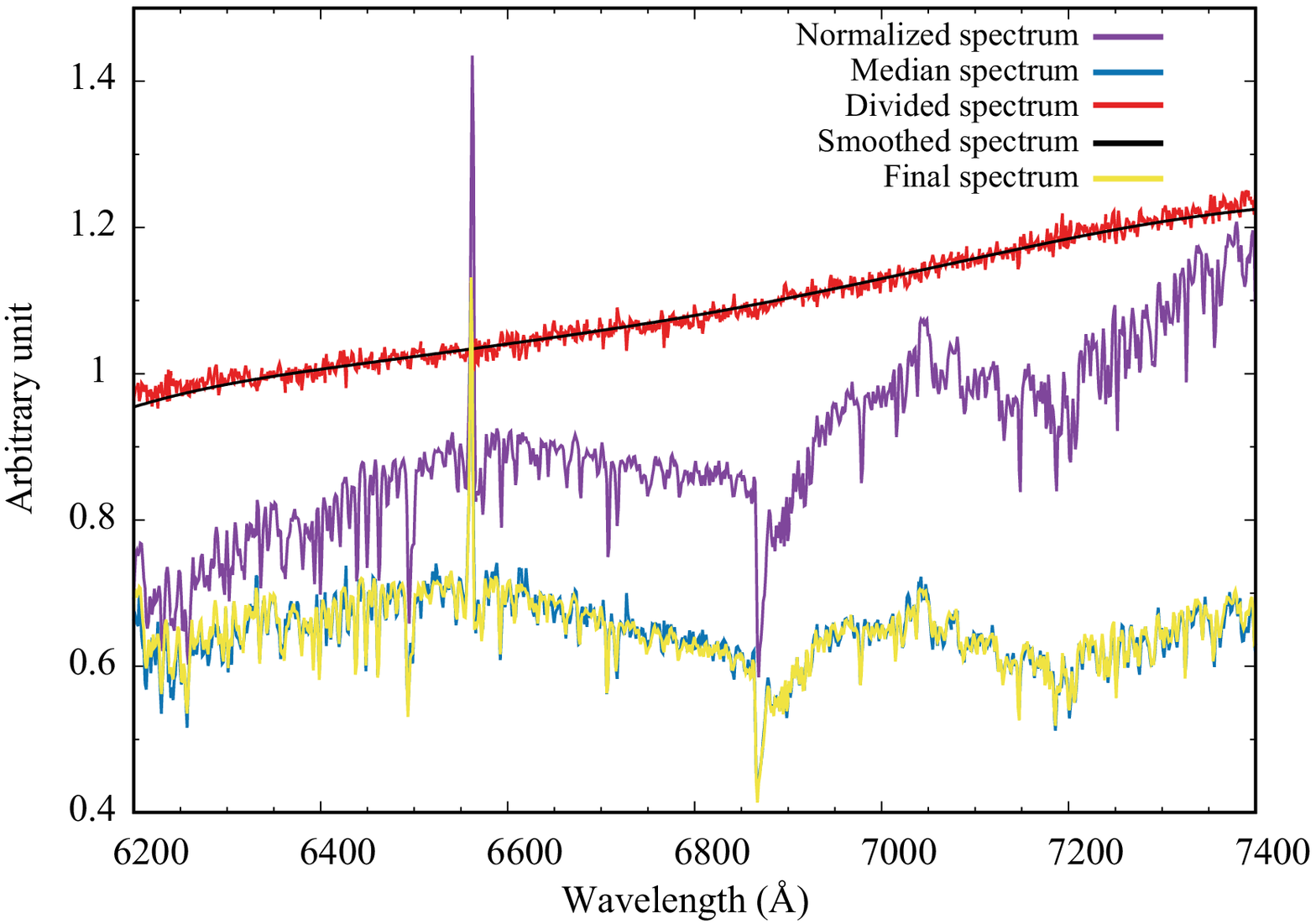}
    \end{centering}
    \caption{Stepwise correction of the fake continuum pattern in each spatial pixel. A purple line is a spectrum of each spatial pixel normalized by a median spectrum of 9,600 spectra (a blue line). A red spectrum represents that the purple spectrum is divided by the blue one. A black line is a Gaussian smoothed spectrum of the red divided one. A yellow line is a final spectrum in which the purple normlized spectrum is divided by the black smoothed one. All spectra are normalized and offset for the presentation purpose.}
    \label{figA3}
\end{figure*}

Figure~\ref{figA3} shows each spectrum in the procedure of correcting the fake continuum pattern in \S~\ref{sec:pca}.

\bibliography{reference}

\end{document}